\documentstyle[pra,aps,epsfig,array,twocolumn]{revtex}
\begin{document}
\def\eps{\epsilon}
\newcommand{\ket}[1]{|  #1 \rangle}
\newcommand{\bra}[1]{ \langle #1  |}
\newcommand{\proj}[1]{\ket{#1}\bra{#1}}
\newcommand{\braket}[2]{\langle #1 | #2 \rangle}
\newcommand{\abs}[1]{ | \, #1 \,  |}
\newcommand{\av}[1]{\langle\,#1\,\rangle}
\newcommand{\asred}{\preceq}
\newcommand{\asredchem}{\leadsto}
\newcommand{\asequ}{\approx}
\newcommand{\exred}{\le}
\newcommand{\exredchem}{\to}
\newcommand{\exequ}{\equiv}
\newcommand{\exequchem}{\rightleftharpoons}
\newcommand{\stored}{\lesssim}
\newcommand{\tr}[2]{{\rm tr}_{\rm \scriptscriptstyle #1}(#2)}
\newcommand{\identity}{\mbox{\boldmath{1} \hspace{-0.33cm}
\boldmath{1} }}
\newcommand{\PT}[2]{(#1)^{\rm T_{#2}}}
\newcommand{\upp}[1]{^{\rm \scriptscriptstyle #1}}
\newcommand{\dnn}[1]{_{\rm \scriptscriptstyle #1}}
\newcommand\bea{\begin{eqnarray}}
\newcommand\eea{\end{eqnarray}}
\newcommand{\beq}{\begin{equation}}
\newcommand{\eeq}{\end{equation}}
\newtheorem{lem}{Lemma}
\newtheorem{theo}{Theorem}
\newtheorem{dfn}{Definition}
\newtheorem{cor}{Corollary}

\twocolumn[\hsize\textwidth\columnwidth\hsize\csname
@twocolumnfalse\endcsname
\title{Remote State Preparation}
\author{Charles H. Bennett$^*$, David P. DiVincenzo$^*$, Peter W.
Shor$^\dag$, John A. Smolin$^*$, Barbara M. Terhal$^*$ and
William
K. Wootters$^\ddag$}
\address{$^*$IBM Research Division, Yorktown Heights, NY 10598,
USA ---
{\tt bennetc, divince, smolin, terhal@watson.ibm.com}}
\address{$^\dag$ AT\&T Labs---Research, Florham Park, NJ 07932, USA
--- {\tt shor@research.att.com}}
\address{$^\ddag$Department of Physics, Williams College,
Williamstown, MA
01267 ---
{\tt William.K.Wootters@williams.edu}}

\date{\today}
\maketitle
\begin{abstract}
Quantum teleportation uses prior entanglement and forward
classical communication to transmit one instance of an unknown
quantum state. Remote state preparation (RSP) has the same goal,
but the sender knows classically what state is to be transmitted.
We show that the asymptotic classical communication cost of RSP
is one bit per qubit---half that of teleportation---and even less
when transmitting part of a known entangled state.  We explore
the tradeoff between entanglement and classical communication
required for RSP, and discuss RSP capacities of general quantum channels.
\end{abstract} 
]

A principal goal of quantum information theory is understanding
the resources necessary and sufficient for intact transmission of
quantum states. In quantum teleportation~\cite{teleport} an
unknown state is transmitted from a sender (``Alice'') to a
receiver (``Bob'') using classical communication and prior
entanglement. Two bits of forward classical communication and one
ebit of entanglement (a maximally entangled pair of qubits) per
teleported qubit are both necessary and sufficient, and neither
resource can be traded off against the other. In remote state
preparation (RSP) the goal is the same---for Bob to end up with a
single specimen of a state---but here Alice starts with complete
classical knowledge of the state.

Pati~\cite{Pati} and Lo~\cite{Lo99} showed that for special
ensembles of states (e.g. qubit states on the equator of the
Bloch sphere) RSP requires less classical communication than
teleportation, but Lo conjectured that for general states the
classical communication costs of the two tasks would be equal.
Here we show that, in the presence of a large amount of prior
entanglement, the asymptotic classical communication cost of RSP
for general states is one bit per qubit, half that of
teleportation. Most of this entanglement is not destroyed, but,
as we will show, can be recovered afterward using {\em backward\/}
classical communication from Bob to Alice, a resource that is
entirely unhelpful for teleportation.

We show that RSP is unlike teleportation in that it exhibits a
nontrivial tradeoff between classical communication and
entanglement, the classical cost of preparing a generic qubit
state ranging from one bit in the high entanglement limit to
infinitely many without prior entanglement (if any finite
classical message, say of $k$ bits, sufficed, Bob could use that
message to make infinitely many copies, determine the state's
amplitudes to more than $k$ bits precision, and thereby violate
causality).

We introduce two new kinds of channel capacity, reflecting a
general quantum channel's asymptotic ability to be used for
remote
state preparation, with or without prior entanglement, and relate
these capacities to the regular quantum and classical capacities
with or without prior entanglement. Finally, we discuss remote
preparation of states entangled between Alice and Bob.

\noindent {\bf RSP in the high-entanglement limit:} To see how a
large amount of shared entanglement enables general states to be
remotely prepared at an asymptotic cost of one bit per qubit, it
is helpful first to consider an exact (non-asymptotic) RSP
protocol for the special ensemble mentioned earlier: equatorial
states. Assume Alice and Bob share a number of singlets, i.e.
pairs of qubits in the state $\ket{\Psi^-}=\ket{01}\!-\!\ket{10}$
(we will often omit the normalization $1/\sqrt{2}$). To remotely
prepare an equatorial state $\ket{\psi}=\ket{0}+e^{i\phi}\ket{1}$,
Alice takes one of her singlets and measures it~\cite{Pati} in the
basis $(\psi, \psi^\perp)$ where $\psi^\perp$ denotes the
antipodal (orthogonal) state to $\psi$. If the outcome is
$\psi^\perp$ she knows (by the properties of the singlet state)
that Bob's remaining half of the singlet is in the desired state
$\psi$. But equally often Alice's outcome is $\psi$, leaving Bob
with $\psi^\perp$, the antipode of the state Alice wished to
prepare. For equatorial states, Bob can correct $\psi^\perp$ to
$\psi$ by applying the Pauli operator $\sigma_z$, a 180 degree
rotation about the $z$ axis. Thus Alice can remotely prepare an
arbitrary equatorial state known to her by measuring a shared
singlet in the basis determined by that state, and sending Bob the
one-bit measurement result, which tells him whether to apply
$\sigma_z$. But for general, non-equatorial states, the corrective
transformation $\psi^\perp\rightarrow\psi$ is antiunitary, and Bob
cannot perform it by any physical means.

Now suppose Alice wishes to remotely prepare a large number of
general qubit states $\psi_1, \psi_2, \ldots,\psi_n$, and that
she and Bob share an unlimited supply of singlets. For each
$j\!=\!1\ldots n$, Alice measures $m=2^{n+\log n}$ of her
singlets
in the basis $\{\psi_j,\psi_j^\perp\}$, and stores the results as
one row of an $n\times m$ table $T$, writing $T(j,k)=1$ for a
success (meaning Bob's half of that singlet is in the desired
state $\psi_j$) and $T(j,k)=0$ for a failure (meaning Bob's half
is in the antipodal state $\psi_j^\perp$). Alice does all this
without telling Bob anything, obtaining a large table of $mn$
independent random zeros and ones. When she is done making all
the
measurements, she looks for a column of all ones, and uses
$n\!+\!\log n$ bits to tell Bob its index. Bob keeps the states
in
the successful column and discards all the others. If (with
probability $o(1)$) no successful column exists, Alice tells Bob
so, then uses $n$ more singlets and $2n$ classical bits to simply
teleport the states to Bob. Thus 1 bit per qubit is
asymptotically sufficient for RSP; it is also necessary~\cite{Lo99} by
causality.

This protocol can be generalized from qubits to states in a
$d$-dimensional Hilbert space, allowing them to be remotely
prepared at an asymptotic classical communication cost of $\log_2 d$
bits per state. Instead of singlets, Alice and Bob use maximally
entangled pairs of
the form $\ket{\Phi^+_d}=\ket{00}+\ket{11}+\ldots
+\ket{(d\!-\!1)(d\!-\!1)}$.
Alice and Bob prearrange $mn$ such states with $m \!\gg\! d^n$ in
an array of
$n$ rows and $m$ columns. For each row $j$, Alice measures her
halves of
the pairs in a basis including $\psi_j^*$, the complex
conjugate of the state she wishes to remotely prepare. If (with
probability $1/d$) her measurement outcome is $\psi_j^*$, Bob's
half of
the entangled pair will be left in the desired state $\psi_j$,
and Alice
enters a 1 in her success/failure table; otherwise she enters a
0.


The high-entanglement RSP protocol described above uses a large
number of ebits, approximately $2^n$ per state sent if $n$ qubits
are transmitted.  But, using back communication, this protocol can
be modified so that only a constant number of ebits are needed per
state transmitted, while still only requiring one classical bit.
To achieve this, we first (following a suggestion of A. Ambainis)
divide the $n$ states to be transmitted into subblocks of size
$s$; $s\rightarrow\infty$ as $n\rightarrow\infty$, but
$2^s/n\rightarrow 0$.  Within each subblock the basic scheme
described above is followed.  But instead of performing a separate
von Neumann measurement on her half of each of the ebits, Alice
does a less intrusive measurement: for each set of $s$ ebits
constituting a column in her table, she performs a two-outcome
incomplete von Neumann measurement.  The ``1'' outcome, obtained
with probability $2^{-s}$, signals that {\em all} Bob's particles
are in the desired state $\Pi_{j=1}^s\ket{\psi_j}$; the ``0''
outcome signals all other possibilities. The joint state remaining
between Alice and Bob when ``0'' is obtained, $\rho_0$, is still
highly entangled, and pure entanglement can be recovered from it
by distillation. From Bob's viewpoint the state $\rho_0$ is mixed,
because he does not know the bases of Alice's measurements.
Averaging over all such bases, the diagonal elements of $\rho_0$
in the generalized Bell basis are:
\begin{equation}
\langle B|\rho_0|B\rangle=\frac{2^s-2}{2^s-1}\delta_{sr}+
\left({1\over 3}\right)^{s-r}\frac{1}{2^s(2^s-1)},\label{diag}
\end{equation}
where $\ket{B}$ is any tensor product of Bell states
$\{\Phi^{\pm}=\ket{00}\pm\ket{11},\Psi^\pm=\ket{01}\pm\ket{10}\}$
containing $r$ instances of $\Phi^+$ and $s\!-\!r$ instances of
the other Bell states.  Alice and Bob collect all these $\rho_0$
states until $s'$ RSPs have been performed, with $s \!\ll\! s' \!
\ll\! n$; at this point they have about $c=s'2^s/s$ copies of
$\rho_0$. They then perform an entanglement distillation
procedure.  After dephasing in the Bell basis (which can be
accomplished by a twirling\cite{BDSW} performed by Alice and Bob),
the state $\rho_0^{\otimes c}$ can be approximated, using a
typical subset, by an equal mixture of $2^{cS}$ different products
of Bell states.  Here the von Neumann entropy of the twirled
$\rho_0$ is $S=s2^{-s}(2+{1\over 2}\log 3)$.  By the
random-hashing technique\cite{BDSW}, $c(s-S)$ pure singlets can be
distilled from this mixture with the help of back communication
from Bob to Alice.  Counting also the one pair consumed when the
successful ``1'' outcome is obtained, the number of ebits consumed
per state transmitted becomes $e_0=1+cS/s'=3+{1\over 2}\log
3\approx 3.79$. This point $(e_0,b=1)$ is labelled {\bf R} in Fig.
1.

\noindent {\bf More restricted protocols and Lo's conjecture:} For
any set of $n$ states to be remotely prepared, the above protocols
are {\em exactly faithful\/}, i.e.\ always work, reproducing
exactly the desired output even for finite $n$, but only {\em
asymptotically efficient\/}, since the expected classical
communication approaches one bit per qubit only in the limit of
large $n$, while for any finite $n$, there is some chance that the
classical communication cost will exceed that required for
teleportation.  We know of no exactly faithful RSP protocol for
finite $n$ that always uses less classical communication than
would be required by teleportation.  In this sense Lo's conjecture
still stands.

In a more restricted setting we can prove Lo's conjecture. Suppose
Alice wants to remotely prepare {\em a single} quantum state
$\psi$ in a $d$-dimensional Hilbert space (for simplicity $d$ is a
power of 2) for Bob.  As in teleportation, we restrict Bob to
performing a unitary transformation on some system in his lab
determined by the classical data he receives from Alice. Also, as
in teleportation, we require that the probability that Alice sends
message $i$ to Bob not depend on the state that she is remotely
preparing. If such a protocol is  exactly faithful, we can show
that it must use at least $2 \log d$ classical bits of
communication from Alice to Bob, as in teleportation. The argument
is as follows. Let $k$ be the number of classical bits that Alice
sends to remotely prepare $\psi$. We will have Bob guess this
data. He infers from the protocol that he will get message $i$
($i=1,\ldots,2^k$) with probability $p_i$ ($\sum_{i=1}^{2^k}
p_i=1$). Thus he flips a coin with bias $p_i$ and he implements
the corresponding unitary transformation in his lab. Since the
protocol only allows him to carry out unitary transformations,
guessing wrong means that instead of getting $\ket{\psi}$ he will
obtain $U \ket{\psi}$ where $U$ is some unitary transformation.
The total probability $p$ of Bob guessing correctly is given by
the sum over $i$ of the probability that Alice sends $i$ and Bob
correctly guesses $i$, which is $\sum_i p_i^2 \geq 2^{-k} \sum_i
p_i=2^{-k}$.
Alice and Bob have thus created a channel ${\cal S}$ which acts
upon the state $\psi$ in Alice's lab and outputs the state
$\rho={\cal S}(\proj{\psi})=p \proj{\psi}+ (1-p) {\cal S}_{\rm
wrong}(\proj{\psi})$ where $p \geq 2^{-k}$. Since Bob used zero
communication to make this state, it must be that \beq f({\cal
S})\equiv\frac{1}{{\rm Vol}(\psi)} \int d \psi\; \bra{\psi}\,{\cal
S}(\proj{\psi})\, \ket{\psi} \leq \frac{1}{d}.
\label{fgreater1overd} \eeq If not, Alice and Bob would have
created a superluminal channel. We can use a result by the
Horodeckis \cite{gentele} which relates $f({\cal S})$ to the
maximally entangled fraction $F({\cal S})\equiv \bra{\Phi^+_d}\,
({\bf 1} \otimes {\cal S})(\ket{\Phi^+_d} \bra{\Phi^+_d})
\,\ket{\Phi^+_d},$ i.e. $f({\cal S})=(F({\cal S}) d+1)/(d+1)$.
Since ${\cal S}$ is the identity operator with probability larger
than or equal to $2^{-k}$ we have $f({\cal S}) \geq (2^{-k}
d+1)/(d+1) > 1/d$ for $k < 2\log d$ in contradiction to
(\ref{fgreater1overd}). Thus in a very restricted
``teleportation'' type of RSP, Lo's conjecture still holds.
Besides being exactly faithful, this restricted protocol is {\em
oblivious}; Bob receives no additional information about $\psi$
other than the state $\psi$ itself. This is due to the fact that
the probability with which Alice sends a classical message does
not depend on the state $\psi$. In the high-entanglement RSP
protocol, by contrast, Bob can gain some additional information
about $\psi$ by measuring the singlets in the unsuccessful columns
instead of recycling them. Perhaps Lo's conjecture holds for all
oblivious, exactly faithful protocols.

For the next two sections we relax the requirement of exact
fidelity, requiring only that protocols be {\em asymptotically
faithful\/}, i.e. for any set of $n$ input states, they should
produce an approximation to the desired output
$\psi_1\otimes\psi_2\otimes \ldots \otimes \psi_n$ whose fidelity
approaches 1 in the limit of large $n$. This definition has the
advantage of allowing RSP to be composed with other asymptotically
faithful processes such as Schumacher compression
\cite{schumacher}.

\noindent {\bf Low-entanglement RSP:} Here we bound the forward
classical communication $b$ needed to remotely prepare qubit
states using entanglement $e<1$ ebit per qubit. To do so, Alice
sends Bob some classical information about the states
$\psi_1...\psi_n$, so as to reduce their posterior von Neumann
entropy from his viewpoint and allow her to teleport them using
$<1$ ebit per qubit. For example, a qubit uniformly distributed
over a circular cap $C_\theta$ of radius $\theta<\pi$ and area
$A(\theta)\!=\!2\pi(1\!-\!\cos\theta)$ centered on the north pole
has von Neumann entropy $S(\theta)=H_2((1\!-\!\cos\theta)/4)<1$
and can be teleported at an asymptotic cost of $2S(\theta)$ bits
and $S(\theta)$ ebits.

First assume the states $\psi_1...\psi_n$ are uniformly
distributed (a restriction we later remove). For each block length
$n$ and cap radius $\theta$, suppose Alice and Bob have agreed on
an $n$ by $m=n(4\pi/A(\theta))^n$ array of random rotations
$R(i,j), i\!=\!1...n,j\!=\!1...m$. Then, given the states
$\psi_1...\psi_n$, Alice constructs a success/failure table where
a success, $T(i,j)=1$, is counted iff the rotated state
$R(i,j)\psi_i$ falls within the standard cap $C_\theta$. As before
she looks for an all-successful column, and uses an expected
$S'(\theta)+o(1)$ bits per state, where
$S'(\theta)=\log_2(4\pi/A(\theta))$, to tell Bob its index $j$.
Finally, she Schumacher compresses the states in the successful
column and teleports them, at an additional asymptotic cost of
$2S(\theta)$ bits and $S(\theta)$ ebits per state, to Bob, who
rotates the them back into their original positions.  If there is
no successful column, Alice teleports the states directly, without
compression; but this happens so rarely as to not increase the
asymptotic entanglement and communication costs, $e=S(\theta)$ and
$b=S'(\theta)+2S(\theta)$.  The $R$ rotations need not actually be
random: for each $n$ and $\theta$, there always exists a
deterministic set of rotations which performs no worse than
average on uniformly distributed $\psi_i$. We use $D(i,j)$ to
denote these deterministic rotations.

To make the protocol work on arbitrary sequences of states, even
ones maliciously chosen to avoid successes with the particular
rotations $\{D(i,j)\}$ Alice and Bob are using, Alice divides the
states into subblocks of size $s\approx \sqrt{n}$, and applies the
above protocol separately to each subblock, but before doing so
applies a set of $s$ random prerotations $r_1,...r_s$ which Bob
removes afterward, to the states in each subblock. Then, even if
the original states $\psi_i$ are awkwardly located, the randomized
states $r_{i \,{\mathrm mod}\, s}\,\psi_i$ will be random within
each subblock. Reusing the prerotations causes the deviations of
the actual mixed-state output from the ideal $\psi_1...\psi_n$ to
be correlated between subblocks, but because of the exponentially
fast convergence of Schumacher compression with increasing
subblock size, the full $n$-fold fidelity still approaches unity
in the limit $n\!\!\rightarrow\!\!\infty$, for any sequence
$\psi_1...\psi_n$ of states to be remotely prepared. Of course
Alice must tell Bob the prerotations $r_1...r_s$ so he can remove
them at the end.  If the prerotations are described with
precision, say, $\sqrt{s}$ bits, the finite-precision errors will
vanish exponentially rapidly, while keeping the communication
overhead sublinear in $n$. \vbox{\vspace{-.16in}
\begin{figure}[htbf]
\epsfxsize=8.1cm \epsfbox{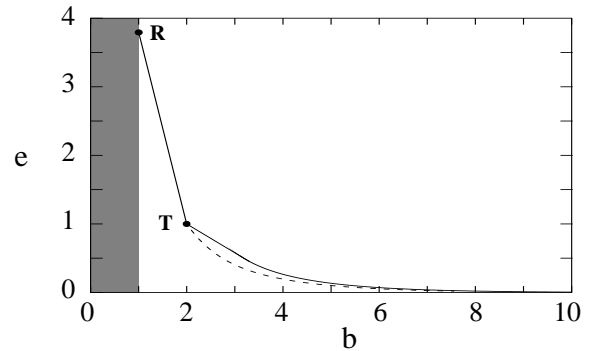}
\smallskip
\caption{Entanglement ($e$) and forward classical communication
($b$) costs of remotely preparing qubit states in various ways,
including teleportation (T), our high-entanglement method with
entanglement recycling (R), and convex combinations (solid line
between T and R). The shaded region $b<1$ is inaccessible because
it would violate causality.  Solid curve below and right of T is
our low-entanglement method and convex combinations with
teleportation. Dashed curve is Devetak-Berger method.}
\vspace{-.10in} \label{rspfig}
\end{figure}
}

Recently Devetak and Berger~\cite{DB01} introduced an improved
protocol which they prove optimal among low-entanglement RSP
methods that use a classical message followed by teleportation to
remotely prepare states uniformly distributed on the Bloch sphere.
Shown as the dashed curve in Figure 1, their method is like our
low-entanglement methods, but instead of the index of the first
successful column, Alice tells Bob the index of the column whose
states, viewed as a finite ensemble, have least entropy.

\noindent {\bf RSP Capacities of Quantum Channels:} Our results
suggest new kinds of capacity for a general noisy quantum channel,
expressing its asymptotic ability to send known states, with or
without the help of prior shared entanglement. For any channel
${\cal N}$ we define the RSP capacity (which might depend on the
dimension $d$ of the Hilbert space ${\cal H}_d$) as
\begin{eqnarray}
& & R^{(d)}({\cal
N})=\lim_{\eps\rightarrow0}\limsup_{m\rightarrow\infty}\;
\{\frac{n\log d}{m}:
\exists_{{\cal D}_{mn}}\forall_{\psi_1,\ldots,\psi_n\in {\cal
H}_d}
\exists_{{\cal E}_{mn}}  \nonumber \\
& & F(\psi_1\otimes \ldots \otimes \psi_n,
{\cal D}_{mn}{\cal N}^{\otimes m}{\cal
E}_{mn})\;>1\!-\!\eps\;\;\},
\end{eqnarray}
where ${\cal E}_{mn}$ denotes a possible block encoder used by
Alice, using $n$ classically described states
$\psi_1,\ldots,\psi_n$ to prepare an input to the quantum channel
${\cal N}^{\otimes m}$ (i.e. $m$ parallel instances of ${\cal
N}$); similarly ${\cal D}_{mn}$ denotes a possible block decoder
used by Bob, mapping the $m$ channel outputs to some approximation
of the state to be remotely prepared; and $F(\psi_1\otimes \ldots
\otimes \psi_n, {\cal D}_{mn}{\cal N}^{\otimes m}{\cal E}_{mn})$
denotes the fidelity of this approximation [the fidelity of a pure
state $\Psi$ under linear map ${\cal M}$ is naturally defined as
$F(\Psi,{\cal M})={\mathrm Tr}\Psi{\cal M}(\Psi)$]. The
entanglement-assisted RSP capacity $R_E({\cal N})$ is defined
similarly, except that the encoder and decoder share unlimited
prior entanglement.

Clearly, for any channel, $R^{(d)}\leq C$, since the classical
capacity $C$ may be viewed as the channel's ability to remotely
prepare classical states (i.e. orthogonal states in some basis).
On the other hand, $R^{(d)} \geq Q$, the quantum capacity, since
the efficiency of transmitting known states must be at least that
of transmitting unknown states.

In the entanglement-assisted setting, we can show that $R_E$ is
independent of $d$ and equal to $C_E$, the channel's
entanglement-assisted classical capacity~\cite{CE}. This follows
from the fact that $\log d$ bits of classical communication are
asymptotically both necessary and sufficient to remotely prepare
a general $d$-dimensional state.

Without entanglement, there are channels for which
$R^{(2)}\!>\!Q$, for example a strongly dephasing qubit channel
with $C\!=\!1$ and $0\!<\!Q\!\ll\!1$. Given any point $(e,b)$ on
the dashed curve in Fig.~1, such a channel can be used $n$ times
to share $\approx Qn$ ebits and another $n$ times to transmit $n$
classical bits, giving $R^{(2)}\geq\min\{Q/2e,1/2b\}$
asymptotically; hence $R^{(2)}/Q\geq1/2e$ for small enough $Q$. On
the other hand, $R^{(d)}\!\!=\!\!0$ for any purely classical
channel (i.e.~one with $Q\!\!=\!\!0$), by causality.

\noindent {\bf Remote Preparation of Entangled States:} Like
teleportation, RSP can be applied not only to pure states, but
also to parts of entangled states. However, unlike teleportation,
RSP requires {\em less\/} classical communication to prepare an
entangled state in ${\cal H}_A\otimes{\cal H}_B$, where ${\cal
H}_A$ remains in Alice's lab, than to prepare a pure state in
${\cal H}_B$. To take an extreme example, the standard maximally
entangled state $\Phi^+_d$ in $d\times d$ dimensions can be
converted into any other maximally entangled state in $d\times d$
dimensions with no classical communication at all, because
maximally entangled states are interconvertible by local unitary
operations of Alice. Suppose more generally that Alice and Bob
share an unlimited supply of ebits, and that Alice wants to
prepare a state $\psi \in {\cal H}_A \otimes {\cal H}_B$, which is
known to her. We assume both Hilbert spaces have dimension $d$; if
necessary the smaller can be extended to make this so. Any state
$\psi\in H_A \otimes H_B$ can be written in Schmidt form as
$\ket{\psi}=\sum_{i=1}^d\sqrt{\lambda_i}\ket{a_i}\otimes\ket{b_i}$,
where some of the $\lambda_i$ may be zero.  We give a
probabilistic procedure by which Alice can convert the standard
state $\Phi^+_d$ into the desired $\psi$ with success probability
$1/d$ if $\psi$ is separable and greater than $1/d$ if $\psi$ is
entangled.

Alice begins by bringing the standard state to the form
$U_A\ket{\Phi^+_d}=\ket{\phi}= \frac{1}{\sqrt{d}}
\sum_{i=1}^d\ket{a_i}\otimes\ket{b_i}$ by means of a local unitary
transformation $U_A$.  She then performs a local filtering
operation on it, which can be described by a
positive-operator-valued measure with two elements, $\Pi_1$
(success) and $\Pi_0$ (failure), the resulting state in each case
being $(\sqrt{\Pi_j}\otimes I)\ket{\phi}$.  Here we take $\Pi_1 =
\frac{1}{\Lambda}\sum_{i=1}^d \lambda_i \ket{a_i}\bra{a_i}$ and
$\Pi_0 = I - \Pi_1$, where $\Lambda=\max\{\lambda_i\}$. Success,
which leaves the system in the desired state $\psi$, occurs with
probability $|(\sqrt{\Pi_1}\otimes I) \ket{\phi}|^2=1/(\Lambda
d)$, which is greater than $1/d$ if $\psi$ is entangled.  This
procedure is exactly faithful and asymptotically efficient in the
sense that for any sequence of states $\psi_1...\psi_n\in{\cal
H}_A\otimes{\cal H}_B$ the expected classical cost is $\sum_j\log
(\Lambda_jd)+O(1)$ bits.

As with unentangled states, causality sets a lower bound on the
classical cost of RSP for entangled states.  The cost of RSP for a
set of states $\psi_1...\psi_n$ must be at least $S(\bar{\rho})
-\frac{1}{n}\sum_{i\!=\!1}^nS(\rho_i)$ bits, where $\rho_i =
\tr{A}{\ket{\psi_i}\bra{\psi_i}}$ and
$\bar{\rho}\!=\!\frac{1}{n}\sum_{i\!=\!1}^n\rho_i$, because the
states could be asymptotically used to encode that much classical
information \cite{capacity}. We are investigating how closely this
bound can be approached.

RSP can be generalized to multiparty
scenarios. For example one may ask whether Alice, using prior
entanglement shared separately with Bob and Charlie, can remotely
prepare an arbitrary tripartite state by sending
$\leq\log d_B$ bits to Bob and $\leq\log d_C$ bits to Charlie.

We thank Andris Ambainis, Igor Devetak and Ashish Thapliyal for
helpful
discussions. CHB, DPD, JAS and BMT acknowledge support from the
US
Army Research Office, grant DAAG55-98-C-0041.

\end{document}